# Flying Drones to Locate Cyber-Attackers in LoRaWAN Metropolitan Networks

Matteo Repetto[1], Enrico Cambiaso[2], Fabio Patrone[3], and Sandro Zappatore[3]

[1] IMATI, CNR, Genoa, Italy
matteo.repetto@ge.imati.cnr.it
[2] IEIIT, CNR, Genoa, Italy
enrico.cambiaso@cnr.it
[3] DITEN, University of Genoa, Genoa, Italy
fabio.patrone01@unige.it

**Abstract**

Today, many critical services and industrial systems rely on wireless networks for interaction with the Internet of Things (IoT), hence becoming vulnerable to a broad number of cyber-threats. While detecting this kind of attacks is not difficult with common cyber-security tools, and even trivial for jamming, finding their origin and identifying culprits is almost impossible today, yet indispensable to stop them, especially when attacks are generated with portable or self-made devices that continuously move around.

To address this open challenge, the FOLLOWME project investigates the feasibility of using Unmanned Aerial Vehicles (UAVs) to locate and even chase attackers during illicit usage of the radio spectrum. The main objective is to develop a cyber-physical security framework that integrates network telemetry with wireless localization. The former triggers alarms in case of anomalies or known attack patterns and provides a coarse-grained indication of the physical area (i.e., the position of affected access gateways), whereas the latter systematically scans such area to identify the exact location of the attacker. The project will specifically address long-range metropolitan area networks and focus on the LoRaWAN protocol, which is the typical scenario for Smart City services.

## 1 Introduction

The ground-breaking evolution in wireless technologies has brought unprecedented opportunities for Cyber-Physical Systems (CPSs), bringing pervasive and ubiquitous connectivity to smart objects and creating the IoT. Beside the evolution of cellular technologies, several short and long-range communication protocols have emerged to fulfill the requirements of different applications, both for licensed and unlicensed bands, boosting new services in almost every industrial domain. As a result, today many critical services and industrial systems rely on wireless networks for communication with people and interaction with the IoT, both in urban and suburban environments. Indeed, IoT sensors and actuators are massively present in applications for Smart Cities, Industry 4.0, Agriculture, Smart Grid, Water and Energy supply.

Despite the evident benefits in terms of infrastructure cost savings and mobility, wireless networks are exposed to more cyber-threats than wired ones due to the unrestricted access to the transmission medium, tampering of devices in public spaces, and poor design/configuration of the IoT. The threat landscape includes brute force Denial of Service (DoS) attacks at the physical layer with continuous, spot, or selective jamming, device tampering, and malware, as well as more elaborated attacks at the data link layer by packet interception, replication, or alteration. Detecting attacks and compromised devices is not difficult with common cyber-security tools, and is even trivial for jamming [20]. However, finding their origin and identifying culprits is very challenging today, yet indispensable to stop them, especially when attacks are generated with portable or self-made devices that continuously move around. This typically requires locating the source of the radio signal, which is possible by measuring the signal strength with directional antennas from multiple positions but cumbersome and time-consuming with physical issues of accessibility and propagation effects to consider.

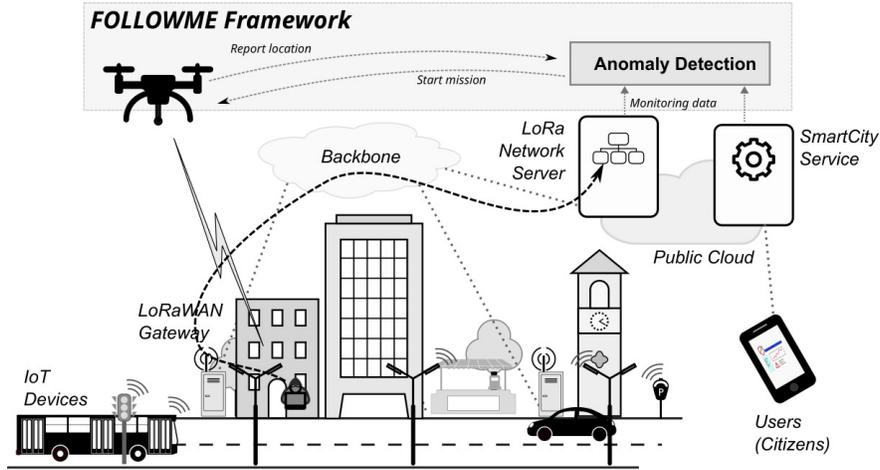

Figure 1: The FOLLOWME concept applied in a Smart City scenario.

The Flying Drones to Locate Cyber-Attackers in LoRaWAN Metropolitan Networks (FOLLOWME) project aims at developing new tools and processes for fighting cyber-crimes against modern digital services by addressing the existing difficulties in locating the origin of attacks on wireless communications. In this respect, it investigates the feasibility of using UAVs, aka flying drones, to locate and even chase attackers during illicit usage of the radio spectrum. The main objective is to develop a cyber-physical security framework that integrates network telemetry with wireless localization, depicted in Figure 1. The former triggers alarms in case of anomalies or known attack patterns and provides a coarse-grained indication of the physical area (i.e., the position of affected access gateways), whereas the latter systematically scans such area to identify the exact location of the attacker.

Without loss of generality, the project focuses on a specific long-range communication protocol, called Long Range (LoRa) Wide Area Network (LoRaWAN), which is the typical scenario for Smart City services [4] and large infrastructures [23]. The scope will include advanced tools for analysis of LoRaWAN traffic and detection of anomalies, localization of LoRa devices, as well as design and field experimentation of a working prototype for the UAV. The results from the project will provide better confidence about the feasibility and performance of the overall FOLLOWME concept, which represents the preliminary step towards extension to other protocols and experimentation with potential users.

The rest of the paper is organized as follows. We review the relevant literature on LoRa threats, Machine Learning (ML)-based detection techniques, and the usage of UAVs for security purposes in Section 2. We describe our investigation approach in Section 3 and how the FOLLOWME approach goes beyond the current State-of-the-Art in Section 4. Finally, we give our conclusion in Section 5.

## 2  Related work

The LoRaWAN protocol is gaining increasing popularity for connecting the IoT in both rural and urban environments and is currently used in several applications. Nevertheless, several vulnerabilities can be exploited to perform different attacks [22]. A non-exhaustive list of such attacks includes eavesdropping, jamming, Man-in-the-Middle (MitM), replay, sinkhole/wormhole, sybil/spoofing, and Distributed DoS (DDoS).

Knowing the vulnerabilities and attacks related to a communication protocol allows researchers to propose and test possible solutions to improve the protocol security that could be part of the following protocol versions. However, when devices using the LoRaWAN protocol are already widespread, updating all of them could be difficult, especially if their number is very high, they are not easily reachable and/or have limited available communication resources, like IoT devices. This highlights the importance of Intrusion Detection Systems (IDS), i.e., systems able to analyze the network behavior and understand if it deviates from the normal status, a possible symptom of ongoing attacks. These systems can be based on different principles, such as ML-based IDS [12], and use different strategies, such as anomaly-based IDS [3]. As the use of ML to identify attacks affecting LoRaWAN is still an open research challenge [19]. Investigating new ML approaches (e.g., eXplainable Artificial Intelligence (AI) (XAI) [25]) to detect cyber-attacks in LoRaWAN networks may lead to relevant results for the research community.

As soon as an intrusion/attack is detected, another important aspect is to locate the attacker. Some of the mentioned attacks involve the transmission of LoRaWAN packets or disturbing signals with the related malicious aim carried on by a malicious device located within the network coverage area. Additional information collected by a LoRaWAN node acting as a gateway, or even a simpler passive receiver, can be exploited for this purpose. Different techniques, such as Time-of-Arrival (ToA), Angle-of-Arrival (AoA), and multilateration have been applied in LoRaWAN networks to locate devices [13]. The main used information is the Received Signal Strength Indicator (RSSI) [11].

Concerning the use of UAVs, multiple purposes have been considered. One of them is communications: aerial communication networks involving UAVs as relay nodes have already been proposed in the literature. They may be composed of different layers with UAVs equipped with different communication interfaces [18]. This includes LoRaWAN communication networks based on UAVs equipped with LoRaWAN interfaces. The UAV task is to collect data from LoRaWAN IoT devices and forward them to the network through other kinds of interfaces, e.g., satellite-based [14].

Another purpose is physical security and surveillance. Most of the solutions proposed in this context involve one or multiple UAVs, remotely or autonomously driven, equipped with cameras to take pictures or videos of the monitored areas [26]. These solutions focus on the surveillance of specific objects and include algorithms that post-process the collected data to detect and track these objects, such as crowd surveillance [17] and traffic surveillance [5]. In [15], the authors consider the possibility to exploit additional information collected by a UAV acting as a LoRaWAN gateway, such as the received packets' Signal-to-Noise Ratio (SNR) or RSSI, to localize and attack the active IoT devices (in case of malicious intentions) or the attacker (in case of defensive intentions).

## 3  The FOLLOWME approach

The FOLLOWME project aims at locating the source of attacks against a LoRaWAN network. We assume that the attack can be detected by inspecting the traffic between the Gateway and the Network or Application Servers, hence without the need to sniff the radio channel. This assumption has negligible impact on the threat model because the Gateway acts as a proxy for the whole LoRaWAN data packet.

The FOLLOWME approach is therefore based on two main components:

(i) an IDS system able to detect attacks against the LoRaWAN network, and

(ii) a UAV which helps to physically locate the attacker, even in dynamic scenarios.

### 3.1 Threat Model

The FOLLOWME concept can address a broad range of attack vectors rooted at LoRaWAN devices, either directly owned by criminals or compromised by malware, tampering, or intrusion. For what concerns the target, we consider any device and application within the LoRaWAN infrastructure (gateways, network servers, application servers) as well as external services and networks.

Relevant attack vectors under consideration include:

- Physical DoS and network attacks, such as eavesdropping [8]: the attacker passively monitors and eavesdrops the data generated by or destined to the end device;

- Jamming [10]: the attacker generates a powerful disturbing signal aimed to disrupt all legitimate communications within its coverage range;

- Man-in-the-Middle [21]: the attacker monitors and intercepts the communications between the end devices and gateways/servers in order to collect data and exploit information, such as the one that is exchanged to secure data transmissions and authenticate the devices;

- Replay [16]: the attacker captures a set of messages exchanged by two nodes and subsequently re-transmits them;

- Sinkhole/Wormhole [9]: the attacker inserts a malicious listening device that intercepts data packets coming from the attacked end device and forwards them to a malicious IoT node that sends these packets to the LoRaWAN gateway pretending to be the real/legitimate device; if the attack is successful, any downlink communications can be re-routed towards the malicious device;

- Sybil/Spoofing [1]: the attacker creates and uses different fake identities to gain the system's trust and violate users' privacy;

- Distributed DoS [24]: the attacker generates a very high number of network requests aimed at saturating the system resources and disabling the system's services.

This fully covers the range of potential attacks where physical identification of the attack source through UAVs makes sense (i.e., those originated by wireless devices). Overall, the project threat model covers the major part of the threat landscape for LoRaWAN.

## 3.2  Detection and localization challenges

The FOLLOWME objective is twofold: on the one hand, the detection of ongoing attacks originated by LoRaWAN devices and, on the other hand, localization of the attack source to stop and prosecute them legally. FOLLOWME pursues a two-tier localization framework with incremental accuracy.

The first *coarse-grained* approximation will be the geo-localization of the area where the attacker is operating by pinpointing the attack to the access gateway serving the malicious node. Even if the identification of such a gateway is usually trivial, the real challenge is the detection of ongoing network attacks. As a matter of fact, LoRaWAN traffic is largely application-specific, hence it does not follow homogeneous patterns, as often happens for Internet traffic. This also means that generic Indicators of Compromise (IoCs) are not available. We will therefore consider anomaly detection methods that leverage XAI to detect deviations from expected network and protocol patterns. The scope includes attacks to both internal LoRaWAN nodes and external elements.

Given the long range of LoRaWAN communications, coarse-grained localization is largely ineffective to find the attacker location with a sufficiently high precision. Fine-grained localization of the malicious device is possible by systematically scanning the radio spectrum in the coverage area. To this purpose, UAVs will be equipped with antennas and the necessary software to process both received signal measures and packets, carrying out the localization task and continuously inspecting the network traffic. Using UAVs allows scanning the relevant area quicker than walking or driving vehicles and reaching inaccessible areas (e.g., pipelines or powerlines across mountains and rural areas), having a better opportunity to work in Line-of-Sight (LoS) than equipment operating along roads.

However, fine-grained localization in radio networks is challenging. First, intermittent traffic (which is the typical case for all attacks but continuous jamming) and moving targets may easily deceive most existing algorithms. Second, the usage of unlicensed radio spectrum and remotely-controlled UAV create interference that may hinder the correct measurements of the malicious bearer. We will therefore select the most promising localization mechanisms and validate them in this specific context.

## 3.3  Scenario

We assume a star-of-stars network topology, which is the typical scenario of many CPSs (e.g., Smart City and Smart Grid). We consider four main components, as shown in Figure 2(a):

1. End devices ($E_i$);
2. Gateways ($G_j$);
3. Network Server ($Ns$);
4. Application Servers ($S_k$).

The gateways receive packets from end devices on LoRa channels and forward them to the Network Server over the Internet. The Network Server terminates the LoRaWAN protocol and delivers data packets to the Application Servers.

The attack scenario is depicted in Figure 2(b). We assume a mobile adversary who moves around the territory and connects to the different gateways following a certain movement path (red dashed trajectory). The adversary initiates attacks according to the threat model previously discussed, even by using multiple vectors. As a result of the traffic analysis performed on the network (in gateways, Network Server, and Application Servers), an alert is triggered at some point by the detection mechanism, including the indication of the gateway $G_i$ to which the attacker is currently connected. At this point, the drone takes off, flights to the macro-area covered by the reported gateway, starts the fine-grained localization, and tracks the movements of the attacker once it has been locked on the target (green dashed trajectory).

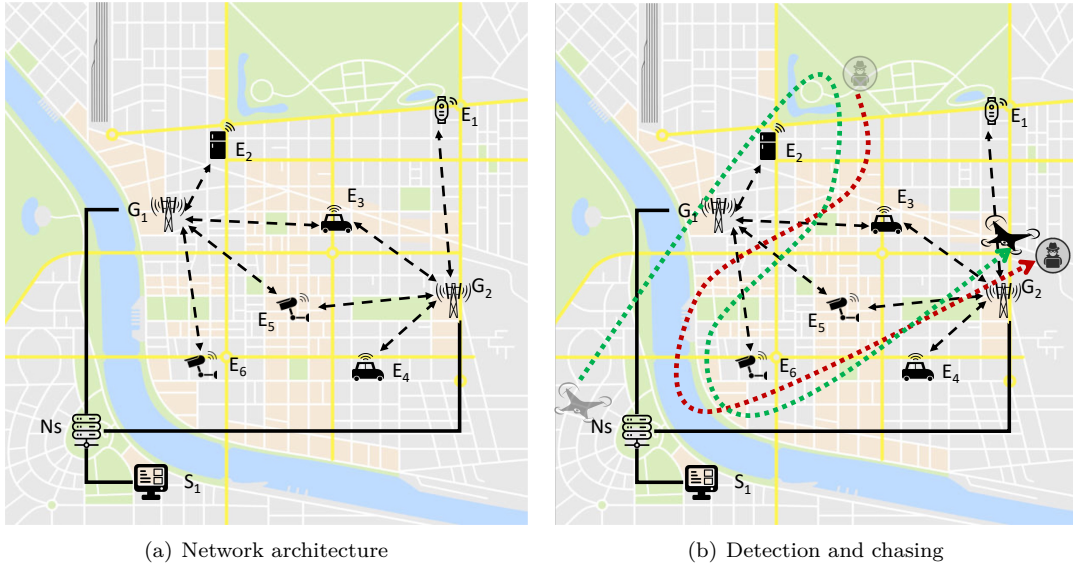

(a) Network architecture  (b) Detection and chasing

Figure 2: Reference scenario for detection and chasing a mobile attacker in a LoRaWAN network.

## 4 Ambition

The FOLLOWME concept is a rare example of mixing cyber and physical tools for fighting crimes against the digital infrastructure. The main innovation behind the project is indeed the concept of cyber-physical security, which coordinates network telemetry with environmental sensing. While research usually focuses on advanced detection methods for cyber-attacks, often proposing automated response and mitigation strategies, it does not preclude that the attack is replicated again in the future. In a similar way, while many applications of UAVs for security and safety purposes have been investigated by using cameras, radars, and other sensing equipment, FOLLOWME explores a new direction related to cyber-security.

FOLLOWME allows eradicating the cause of an attack rather than only mitigating it. It is worth noting that FOLLOWME aims at locating physical people instead of activist groups or avatars that cannot be concretely prosecuted for their crimes.

In the context of radio networks, physical attacks are often easier to implement than cyber ones, but most challenging to mitigate. Jamming is a typical threat that is trivial to detect but almost impossible to mitigate by using only cyber-tools. As soon as critical infrastructures (e.g., energy and water supply) increasingly rely on wireless sensors and smart meters for their correct operation, the FOLLOWME approach becomes one effective way to overcome the limitations of cyber-appliances in fighting this kind of physical threats and ensure service continuity. Moreover, the project implementation is expected to bring additional use cases and applications beside LoRaWAN networks, hence broadening the scope.

The implementation of the FOLLOWME concept aims at developing a complete detection and response framework. It builds on two main assumptions: i) the ability to effectively detect cyber-attacks and ii) the possibility to link their source to physical people. Therefore, FOLLOWME entails the following main activities:

- attack detection;
- attack localization and tracking;
- incident response.

## 4.1 Attack detection

The attack detection objective is twofold. On the one hand, the selection of cyber-attacks to be executed and investigated. The investigation of novel detection algorithms implies the identification and replication of feasible attacks through a LoRaWAN network. The scope is not limited to known attacks already described in the literature, but also entails the discovery of additional vulnerabilities of the LoRaWAN protocol when experimenting with new conditions and illicit usage patterns. An interesting area is the study of possible covert channels that can be realized in packet headers or, more likely, by modulating packet transmission.

On the other hand, an IDS will be designed and developed. The scope entails the detection of malicious/compromised nodes. The threat model includes attacks to the LoRaWAN network itself as well as external assets (e.g., botnets to perform DDoS and amplification attacks against Internet servers). FOLLOWME investigates the usage of AI in the detection process, addressing cyber-attacks that are not currently recognizable in a satisfactory way by traditional anomaly detection mechanisms (i.e., slow DoS [6], hidden tunnels [2], covert channels [7]). Starting from the available information that can be collected in the specific use case environment, the most suitable detection strategies will be defined and implemented, based on the considered threats, by focusing on reducing detection times (e.g. through the use of XAI rules able to reliably and quickly characterize anomalies). This approach will support the selection of new IoCs that can detect attacks without the need for long, time-consuming, and critical training operations.

## 4.2 Localization and tracking

This activity concerns the design and development of the localization and tracking technique. As a consequence of the identification of an ongoing attack and its rough area by the FOLLOWME IDS, this task will focus on the localization of the attacker by monitoring the environment from the drone, collecting valuable information (e.g., RSSI values and packet reception times), combining and analyzing them to accurately infer the position of the attacker through the use of ad-hoc implemented methodologies.

Radio localization is a new field of application for drones, where previous experience is largely missing. Improvement of knowledge includes the following aspects:

(i) precision of the localization algorithm with radio measures taken by a flying UAV (eventually following the target);

(ii) feasibility and possible countermeasures to local external signals, such as the radio control of UAVs, in the same Industrial, Scientific and Medical (ISM) band as the attack signal to be located that risks to saturate the receiving antenna;

(iii) robustness of the localization algorithm in mixed indoor/outdoor applications (the target may be located in public or private buildings).

Here the main research challenge is represented by accurate and robust geo-localization in case of intermittent bearer and interference in unlicensed spectrum bands.

One common drawback of every UAV-based application is the limited duration of batteries, which limits the operational range in both time and space. The presence of additional hardware on-board further exacerbates this issue, further limiting the battery lifetime. The development of a prototype will provide a more precise understanding of the size and weight of the localization hardware. The prototype will also provide more confidence about the most appropriate business proposition, namely whether to develop a standalone localization pack bearable by different drones or integrate the solution into a dedicated drone.

### 4.3 Incident Response

The major ambition for FOLLOWME is the integration of cyber and physical means to create an advanced Incident Response (IR) tool. This includes the definition of a system architecture structure that combines the features of its two components: the IDS and the UAV localizer. The challenge is the definition of new incident response procedures able to combine cyber-threat detection systems with physical localization and tracking activities, aimed to provide additional information on the attack and localize the adversary on a Metropolitan Area Network in the context of cyber-physical security.

A particular emphasis will be posed on defining the interactions and information exchanged between these two elements in order to define when the UAV action has to be triggered and how its operations will be coordinated and data will be collected when the UAV is flying. In this respect, the IDS-to-UAV (and vice versa) communications will be defined by focusing on reducing the response times. The procedures to keep contact with flying drones will be considered as well. An information dashboard will also be implemented to provide useful information to the UAV operator. It will support him/her while steering the UAV to reach the location of the adversary and track him/her by also including visualization of the attack source location on the map. Integration with an open-source tool for IR will be considered as well to raise the Technology Readiness Level (TRL) of the entire solution and demonstrate how it can be integrated into existing practice.

### 4.4 Applications

The natural exploitation of the FOLLOWME outcomes is for the investigation of cyber-physical attacks in LoRaWAN networks. Law enforcement is the most likely user of FOLLOWME UAVs to locate cyber-criminals, triggered by warning either from internal staff or private Security Operation Centers (SOCs). The whole system may also be operated by private organizations to investigate attacks on their own or third-party infrastructures. It is worth noting that the FOLLOWME system can be easily extended to other wireless technology, including 4/5/6G networks. Beneficiaries of FOLLOWME are operators of IoT networks, including industries and municipalities, and final users of services built on such infrastructures, mostly citizens.

The typical insecurity posture of IoT devices, their placement in public spaces, and the usage of wireless links in unlicensed spectrum bands are likely to make IoT networks a popular target for cyber-criminals. Many Smart City infrastructures are indeed building on LoRa to interconnect both fixed and mobile devices. The impact of new methods and tools for detecting

LoRaWAN attacks scales to many civil applications already available and, most of all, planned to be deployed in the next years in dozens of cities and involving millions of citizens.

The FOLLOWME technology represents the basis for more innovative and effective ways of chasing cyber-criminals in case of attacks on critical infrastructures, by running fleets of autonomous UAVs that patrol broader areas than what is possible with a single drone. Beside the integrated FOLLOWME platform, individual technologies have high impact potential in orthogonal domains. Detection of cyber-attacks on LoRaWAN networks can be integrated into more complex cyber appliances. The raw detection algorithms can be integrated into existing SOC and IR tools. Mobile localization with UAVs can instead be used to improve identification for emergency and rescue services.

## 4.5 Experimental results

After finding proper answers to the research questions, the last activity concerns the design, implementation, and validation of a working prototype for demonstration and validation. This includes at least the following elements:

1. geo-localization of the attack source on a geographical map with an accuracy level indication (e.g., large circle first with coarse-grained information which narrows down as soon as fine-grained localization is available);

2. onboarding of geo-localization software on Raspberry Pi or similar small-factor device equipped with LoRa antenna and battery pack;

3. installation of the geo-localization equipment on UAV.

The prototype will provide more confidence about the feasibility of the proposed approach and will effectively sustain the result exploitation with tangible videos in addition to scientific papers.

Evaluation of the tools and mechanisms will be carried out in an experimental environment. The testbed will include multiple LoRaWAN nodes deployed in at least 3 locations around Genoa selected to ensure connectivity with one or two gateways, depending on the types of attacks that will be implemented. End nodes and gateways will be implemented with a mix of Arduino Uno, Arduino MKR WAN 1300/1310, Raspberry Pi equipped with LoRa Hat radio module, and CatWAN USB stick, depending on which combination best suits attack emulation. The Network Server will be implemented by using the software ChirpStack, an open-source framework largely used and suitable to be integrated with the FOLLOWME IDS.

The UAV will be equipped with lightweight and portable hardware components, such as antennas and microcontrollers, to allow attaching them to a commercial drone and allow the UAV to directly collect and analyze network packets.

Finally, validation will encompass the replication of the most appropriate attacks for the reference scenario. A list of potential use cases involving LoRaWAN will be derived to motivate the selection of attack methodologies according to likely targets and criminal intents. Known vulnerabilities of the LoRaWAN protocol and its implementation will also be considered, according to what is reported in the literature and public vulnerability databases (e.g., the Common Vulnerability and Exposure program).

Evaluation of the FOLLOWME system will entail the following aspects, that will be carried out in the experimental testbed:

- functional validation of the interactions among the involved components;

Flying Drones to Locate Cyber-Attackers                              Repetto, Cambiaso, Patrone, Zappatore- assessment of the achievable performance of the developed attack detection and attacker localization techniques;

- demonstration of the overall solution.

## 5 Conclusion

The FOLLOWME project pursues an overarching approach to detect and respond to attacks in CPS environments. The main objective is the eradication of the root cause of attacks rather than mitigating its effects, demotivating attackers, or redirecting them towards other targets. This approach largely improves current practice mainly based on plain mitigation without fighting back, hence contributing to make CPS more resilient to attacks. FOLLOWME makes IoT systems more secure by addressing both the insecurity posture of such devices and the lack of control over the transmission media. It supports humans in physical investigations that would normally require hours or days to locate the source of an attack, especially in case of moving malicious nodes.

The potential impact of the FOLLOWME concept goes beyond CPSs. Most modern digital services are built by chaining heterogeneous components and resources from different domains, smart city applications being no exception. FOLLOWME contributes to the overall security of digital service chains by covering one of the weakest links: wireless networks. This ultimately results in better disaster preparedness against possible disruptions, attacks, and cascading effects. While the unavailability of data from CPSs may severely impact the availability of critical services, FOLLOWME provides the necessary technology and tools to promptly react in case of disruption, minimizing the extension of disruptions and mitigating cascading effects on the rest of the chain.

Beyond the innovative approach, scientific advances are expected in the detection of cyber-attacks, as well as in radio localization, by leveraging the scanning capability of a mobile probe. FOLLOWME addresses a broader scope than existing cyber-defence appliances, largely ineffective for stopping physical threats like jamming and tampering.

Further extensions may include the possibility to take images or videos from the UAV to be used as unchallengeable evidence of the presence and activity of the accused in court. An additional interesting research issue concerns the autonomous driving of UAVs to systematically scan the suspicious area and progressively get closer to the target.

## Acknowledgment

The work has been carried out in the context of the project "FOLLOWME - Flying drOnes to Locate cyber-attackers in LOraWan Metropolitan nEtworks", financed by MUR minister of Italy, under the call "PRIN: PROGETTI DI RICERCA DI RILEVANTE INTERESSE NAZIONALE - Bando 2022 PNRR", ERC field PE7_8, Project code P202245HZF, Project CUP B53D23023810001.## References

[1] Hezam Akram Abdul-Ghani, Dimitri Konstantas, and Mohammed Mahyoub. A comprehensive IoT attacks survey based on a building-blocked reference model. *International Journal of Advanced Computer Science and Applications*, 9(3):355–373, 2018.10